\documentstyle[prl,preprint,aps,epsf]{revtex}

\begin{document}
\bibliographystyle{plain}
\title{Asymptotic Quasinormal Frequencies of Brane-Localized Black Hole} 
\author{ 
D. K. Park\footnote{Email:dkpark@hep.kyungnam.ac.kr 
}}
\address{Department of Physics, Kyungnam University,
Masan, 631-701, Korea.}
\date{\today}
\maketitle

\begin{abstract}
The asymptotic quasinormal frequencies of the brane-localized 
$(4+n)$-dimensional black hole are computed. Since the induced metric on the 
brane is not an exact vacuum solution of the Einstein equation defined on
the brane, the real parts of the quasinormal frequencies 
$ \omega$ do not approach
to the well-known value $T_H \ln 3$ but approach to $T_H \ln k_n$, where 
$k_n$ is a number dependent on the extra dimensions. For the scalar field
perturbation $Re(\omega / T_H) = \ln 3$ is reproduced when $n = 0$. For
$n \neq 0$, however, $Re(\omega / T_H)$ is smaller than $\ln 3$. 
It is shown also that when $n > 4$, $Im(\omega / T_H)$ vanishes
in the scalar field perturbation. For the 
gravitational perturbation it is shown that $Re(\omega / T_H) = \ln 3$ is 
reproduced when $n = 0$ and $n = 4$. For different $n$, however, 
$Re(\omega / T_H)$ is smaller than $\ln 3$. When $n = \infty$, for example,
$Re(\omega / T_H)$ approaches to $\ln (1 + 2 \cos \sqrt{5} \pi) \approx 0.906$.
Unlike the scalar field perturbation $Im(\omega / T_H)$ 
does not vanish regradless
of the number of extra dimensions.
\end{abstract}

\newpage
The stability problem of the black holes, when perturbed 
by the external fields, 
is a long-standing issue in the context of the general 
relativity\cite{regge57,vish70-1,edel70,vish70-2}. It is well-known that 
all perturbations are radiated away, which is characterized by the 
quasinormal modes\cite{press71}. These quasinormal modes are defined
as solutions of the perturbation wave equation, belonging to 
complex-characteristic frequencies and satisfying the boundary conditions
for the purely outgoing waves at infinity and purely ingoing waves at 
the horizon, {\it i.e.}
\begin{eqnarray}
\label{def1}
& &\Psi \sim e^{i \omega z} \hspace{1.0cm} \mbox{as} \hspace{1.0cm}
z \rightarrow -\infty
                                \\ \nonumber
& & \Psi \sim e^{-i \omega z} \hspace{1.0cm} \mbox{as} \hspace{1.0cm}
z \rightarrow \infty
\end{eqnarray}
where $z$ is an appropriate ``tortoise'' coordinate and the time dependence
of the fields is taken as $e^{i \omega t}$. 

In our notation the quasinormal
frequencies $\omega$ should satisfy $Im (\omega) \geq 0$.
As a consequence, the quasinormal modes diverge exponentially at both
boundaries. This makes it extremely difficult to determine the quasinormal
frequencies numerically. This is a main reason why only very few 
frequencies with moderate imaginary parts were known\cite{chandra83}.

About two decades ago Leaver\cite{leaver85} found a 
possibility to compute the quasinormal 
frequencies without having to deal with the corresponding quasinormal 
modes numerically. Instead of the solutions he used the recursion relation
which provides an infinite continued fraction. However, this method has a 
technically problem of convergence when $Im (\omega) >> Re (\omega$). 
This defect of 
the continued fraction method was mostly removed by Nollert\cite{noll93}
by computing the ``remaining'' infinite continued fraction.

For the case of the $4d$ Schwarzschild black hole Nollert showed using his
improved numerical method that the asymptotic quasinormal frequencies
for the scalar field and gravitational perturbations
become 
\begin{equation}
\label{nollert1}
\omega = \frac{\bar{n} + 1/2}{2} i + 0.0874247
\hspace{2.0cm} (\bar{n} = 0, 1, 2, \cdots)
\end{equation}
with, for simplicity,  assuming $r_h = 1$ where $r_H$ is an horizon radius.
This was confirmed by Andersson\cite{ander93}  
by the phase integral method, which is an improved WKB-type technique.

After few years Hod\cite{hod98} claimed surprisingly that the numerical
number in Eq.(\ref{nollert1}) is identified as
\begin{equation}
\label{identify1}
0.0874247 \rightarrow \frac{\ln 3}{4 \pi} = T_H \ln 3
\end{equation}
where $T_H$ is an Hawking temperature.
This identification and the Bohr's correspondence principle naturally
imply that the minimal quantum area is $4 \ln 3$, one of the values
$4 \ln k$ suggested in Ref.\cite{beken95}. This is intriguing from the 
loop quantum gravity\cite{thie01} point of view because it suggests that
the gauge group should be $SO(3)$ rather $SU(2)$\cite{drey03}. Subsequently,
the identification (\ref{identify1}) was analytically shown in 
Ref.\cite{motl02,motl03}. Especially in Ref.\cite{motl03} the authors 
transformed the boundary condition of the quasinormal modes at the horizon
into the monodromy in the complex plane of the radial coordinate. 
We will use this method 
to compute the asymptotic quasinormal frequencies of the brane-localized
$(4+n)$-dimensional Schwarzschild black hole. 
For the quasinormal modes of the other asymptotically flat\footnote{See
also Ref.\cite{horo00,cardo01} for asymptotically non-flat black 
holes} black holes see Ref.\cite{berti03-1,berti03-2,musi04,hod05} and
references therein.

Recently, much attention is paid to the higher-dimensional black holes.
Besides its own theoretical interest the main motivation of it seems to
be the emergence of the $TeV$ scale gravity arising in the 
brane-world scenarios\cite{ark98-1,anto98,rs99-1,rs99-2}, which opens the 
possibility to make a tiny black holes factory in the future 
high-energy colliders such as 
LHC\cite{gidd02-1,dimo01-1,eard02-1,stoj04,card05}. In this reason the 
absorption and emission problems of the higher-dimensional black holes
were extensively explored recently\cite{harris03-1,jung05-1,jung05-2,duffy05}.

The lower quasinormal frequencies for the brane-localized $5$-dimensional
rotating black holes were recently computed numerically\cite{ber03-2}.
In this letter we would 
like to go further for the study of the brane-world black holes by examining
the asymptotic quasinormal frequencies of the brane-localized Schwarzschild 
black holes. Since the metric of the brane-localized Schwarzschild black hole
is induced by the higher-dimensional bulk metric, it is not a vacuum solution of
the Einstein equation defined on the brane. We will show that this fact makes
the real part of the asymptotic quasinormal frequencies not to be the 
well-known value $T_H \ln 3$ but to be $T_H \ln k_n$, where $T_H$ is an
Hawking temperature of the higher-dimensional black hole and $k_n$ is a number
dependent on the extra dimensions.

We start with the $(4 + n)$-dimensional Schwarzschild black hole whose 
metric is given by\cite{tang63,myers86}
\begin{equation}
\label{metric1}
ds_B^2 = - h(r) dt^2 + h^{-1}(r) dr^2 + r^2 d \Omega_{n+2}^2
\end{equation}
where 
\begin{eqnarray}
\label{bozo1}
& & 
h(r) = 1 - \left(\frac{r_H}{r}\right)^{n+1}
                                              \\  \nonumber
& &d \Omega_{n+2}^2 = d \theta_1^2 + \sin^2 \theta_1
\left[d \theta_2^2 + \sin^2 \theta_2 
      \left\{ d \theta_3^2 + \cdots + \sin^2 \theta_n
             \left(d \theta_{n+1}^2 + \sin^2 \theta_{n+1} d \phi^2
                              \right) \cdots \right\} \right]
\end{eqnarray}
and $r_H$ is an horizon radius. In Eq.(\ref{metric1}) the subscript
``B'' stands for ``bulk''.
For the consideration of the brane-localized black hole we should consider
the $4d$-induced metric of $ds_B^2$, which is 
\begin{equation}
\label{metric2}
ds_4^2 = - h(r) dt^2 + h^{-1}(r) dr^2 + r^2 (d \theta^2 + \sin^2 \theta
d\phi^2).
\end{equation}

Now, it is easy to show that the equation $\Box \Phi = 0$, which governs
the scalar field perturbation leads to the following radial equation:
\begin{equation}
\label{radial1}
\frac{h(r)}{r^2} \frac{d}{d r}
\left[ r^2 h(r) \frac{d R}{d r} \right] + 
\left[ \omega^2 - h(r) \frac{\ell (\ell + 1)}{r^2} \right] R = 0.
\end{equation}
When deriving Eq.(\ref{radial1}), we used the factorization condition
$\Phi = e^{i \omega t} R(r) Y_{\ell, m} (\theta, \phi)$. Defining 
$R = \Psi / r$, one can show that Eq.(\ref{radial1}) reduces to the 
following Schr\"{o}dinger-like equation
\begin{equation}
\label{schro1}
\left( h(r) \frac{d}{d r} \right)^2 \Psi + \left[\omega^2 - V_S (r)\right]
\Psi = 0
\end{equation}
where $V_S (r)$ is an effective potential of the scalar field perturbation, 
which is given by
\begin{equation}
\label{ef-poten1}
V_S (r) = h(r) \left( \frac{\ell (\ell + 1)}{r^2} + \frac{1}{r}
\frac{d h}{ d r} \right).
\end{equation}

Now, we would like to derive the wave equation which governs the gravitational
perturbation. In order to derive it we should change the metric itself,
{\it say} $ds_4^2 \rightarrow d\tilde{s}_4^2 = ds_4^2 + \delta s^2$ where
\begin{equation}
\label{metric3}
\delta s^2 = \left[H_0(r) dt d\phi + H_1(r) dr d\phi\right] e^{i \omega t}
\sin \theta \frac{d P_{\ell}}{d \theta} (\cos \theta)
\end{equation}
with assuming $H_0, H_1 << 1$ for the linearization\footnote{The metric change
in Eq.(\ref{metric3}) corresponds to the odd-parity gravitational 
perturbation. For detail see Ref.\cite{vish70-2}}. Then, it is 
straightforward to derive the Ricci tensor $\tilde{R}_{\mu \nu}$ and
curvature scalar $\tilde{R}$. 

It is worthwhile noting that since the metric
$ds_4^2$ in Eq.(\ref{metric2}) is an induced one from the higher-dimensional
bulk metric, it is not an exact vacuum solution of $4d$ Einstein equation. 
Thus, it may 
satisfy the non-vacuum Einstein equation ${\cal E}_{\mu \nu} = T_{\mu \nu}$
where ${\cal E}_{\mu \nu} = \tilde{R}_{\mu \nu} - g_{\mu \nu} \tilde{R} / 2$
and $T_{\mu \nu}$ is an energy-momentum tensor.
Since the perturbation in general changes both the geometry of the spacetime
and the energy-momentum tensor, adding $\delta s^2$ to $ds_4^2$ should makes 
the Einstein equation as 
${\cal E}_{\mu \nu} + \delta {\cal E}_{\mu \nu} = T_{\mu \nu} + 
\delta T_{\mu \nu}$ where $\delta {\cal E}_{\mu \nu}$ and 
$\delta T_{\mu \nu}$ are order of $H_0$ or $H_1$. Although we can not 
compute $\delta T_{\mu \nu}$ by conventional method because we do not know
the matter source nature, we can compute it as following. Firstly, we note
that the nonzero components of $\delta T_{\mu \nu}$ are only 
$\delta T_{t \phi}$, $\delta T_{r \phi}$ and $\delta T_{\theta \phi}$.
This can be easily conjectured by computing the Einstein tensor. One
constraint is a covariant conservation of the energy-momentum tensor
$\left(T^{\mu \nu} + \delta T^{\mu \nu} \right)_{; \mu} = 0$. When 
$\nu = t$, $r$ and $\theta$, this is automatically satisfied. Thus, this 
conservation law generates one constraint. Second constraint comes from
the fact that the effective potential which will be derived later should
coincide with the well-known Regge-Wheeler potential when $n=0$ limit. The
final constraint comes from the fact that the $(t, \phi)$ component of the 
Einstein equation should be derived from $(r, \phi)$ and $(\theta, \phi)$
components. These constraints uniquely determine $\delta T_{\mu \nu}$ and 
the final expressions of the Einstein equation are
\begin{eqnarray}
\label{linear1}
& &-\frac{1}{4} h(r) \frac{d^2 H_0}{d r^2} + 
\left[ \frac{\ell (\ell + 1) - 2}{4 r^2} + \frac{1}{2 r^2} h(r) + 
\frac{1}{2 r} \frac{d h}{d r} + \frac{1}{4} \frac{d^2h}{d r^2} \right] H_0
                                                              \\  \nonumber
& & \hspace{2.0cm}
+ \frac{i \omega}{4} h(r) \left(\frac{d H_1}{d r} + \frac{2}{r} H_1\right) = 
-\frac{1}{4 i \omega} h^2(r)
\left(\frac{2}{r^2} \frac{d h}{d r} + \frac{4}{r} \frac{d^2 h}{d r^2}
+ \frac{d^3 h}{d r^3} \right) H_1
                                                    \\   \nonumber
& &-\frac{i \omega}{4} h^{-1}(r) 
\left(\frac{d H_0}{d r} - \frac{2}{r} H_0 \right) + 
\left[ \frac{\ell (\ell + 1) - 2}{4 r^2} - \frac{\omega^2}{4} h^{-1}(r) +
      \frac{1}{2 r} \frac{d h}{d r} + \frac{1}{4} \frac{d^2h}{d r^2} \right]
H_1 = 0
                                                    \\    \nonumber
& &i \omega h^{-1}(r) H_0 - h(r) \frac{d H_1}{d r} - \frac{d h}{d r} H_1 = 0.
\end{eqnarray}
Eliminating $H_0$ in Eq.(\ref{linear1}) appropriately, one can derive a 
second-order differential equation for solely $H_1$ in the form
\begin{eqnarray}
\label{linear2}
& &\frac{d^2 H_1}{d r^2} + 
\left( 3 h^{-1}(r) \frac{d h}{d r} - \frac{2}{r} \right) \frac{d H_1}{d r}
                                                     \\   \nonumber
& & \hspace{1.0cm}
+ \left[ \left( \omega^2 + \left(\frac{d h}{d r} \right)^2 \right) h^{-2}(r)
        - \left(\frac{\ell (\ell + 1) - 2}{r^2} + \frac{4}{r} \frac{d h}{d r}
        \right) h^{-1}(r) \right] H_1 = 0.
\end{eqnarray}
Defining $H_1 \equiv r h^{-1}(r) \Psi(r)$, we can derive the following 
Schr\"{o}dinger-like equation:
\begin{equation}
\label{schro2}
\left(h(r) \frac{d}{d r}\right)^2 \Psi + 
\left[\omega^2 - V_G(r) \right] \Psi = 0
\end{equation}
where $V_G(r)$ is an effective potential of the gravitational perturbation,
which is given by
\begin{equation}
\label{ef-poten2}
V_G(r) = h(r)
\left( \frac{\ell (\ell + 1)}{r^2} + \frac{1}{r} \frac{d h}{d r} - 
\frac{2}{r^2} \left(1 - h(r) \right) + \frac{d^2 h}{d r^2} \right).
\end{equation}
Using $d h(r) / d r = (n+1) (1 - h(r)) / r$, one can express the effective
potentials $V_S(r)$ and $V_G(r)$ as following 
\begin{equation}
\label{ef-poten3}
V_{\mbox{eff}} (r) = h(r) \left[ \frac{\ell (\ell + 1)}{r^2} + 
\frac{\sigma_n}{r^2}
           \left( 1 - h(r) \right) \right]
\end{equation}
where
\begin{eqnarray}
\label{ef-poten4}
\sigma_n = \left\{ \begin{array}{ll}
                   n+1  & \mbox{for scalar field perturbation}  \\
          - (n+1)^2 - 2  & \hspace{1.0cm}\mbox{for gravitational perturbation}
                  \end{array}
          \right. 
\end{eqnarray}
Thus we can treat the scalar and gravitational perturbation in an unified 
way.

Now, we define a tortoise coordinate $z$ as 
\begin{equation}
\label{tortoise}
\frac{d}{d z} \equiv h(r) \frac{d}{d r} =
\left[1 - \left(\frac{r_H}{r}\right)^{n+1} \right] \frac{d}{d r}.
\end{equation}
Integrating Eq.(\ref{tortoise}), we can make an explicit expression of $z$
in the form
\begin{equation}
\label{tortoise2}
z = r + \frac{r_H}{n+1} \sum_{j=0}^{n} e^{i 2 \pi j / (n+1)}
\ln \left[1 - \frac{r}{r_H} e^{-i 2 \pi j / (n+1)} \right].
\end{equation}
When deriving Eq.(\ref{tortoise2}), we fixed the integration constant by 
imposing that $r=0$ corresponds to $z=0$, which is crucial\cite{motl03} for the 
calculation of the asymptotic quasinormal frequencies. From this tortoise
coordinate and the boundary condition (\ref{def1}) it is straightforward to 
compute the monodromy around the 
singularity $r=1$
\begin{equation}
\label{monod1}
{\cal M}(1) = e^{\omega / T_H}
\end{equation}
where $T_H$ is an Hawking temperature of the $(4+n)$-dimensional 
black hole, {\i.e.} $T_H = (n+1) / 4 \pi r_H$. Following Ref.\cite{motl03}
we will
compute another expression of the monodromy by following a contour given 
in Fig. 3 of Ref.\cite{motl03}  and 
using an analytic continuation. Equating these two expressions enables us to
calculate the 
asymptotic quasinormal frequencies. For the following calculation it is 
important to note that near the naked singularity $r \sim 0$ the 
tortoise coordinate $z$ is proportional to $r^{n+2}$ as
\begin{equation}
\label{tortoise3}
z \sim - \frac{r_H}{n+2} \left(\frac{r}{r_H}\right)^{n+2}.
\end{equation}

Adopting a method used in Ref.\cite{motl03}, we can now compute the asymptotic 
quasinormal frequencies. The explicit expression of the effective potential
given in Eq.(\ref{ef-poten3}) is 
\begin{equation}
\label{ef-poten5}
V_{\mbox{eff}}(r) = \left[1 - \left(\frac{r_H}{r}\right)^{n+1} \right]
\left[\frac{\ell (\ell + 1)}{r^2} + \frac{\sigma_n}{r^2}
\left(\frac{r_H}{r}\right)^{n+1} \right].
\end{equation}
Near the naked singularity $r=0$, therefore, the effective potential
can be written by its dominant term
\begin{equation}
\label{ef-poten6}
V_{\mbox{eff}}(r) \sim - \frac{\sigma_n r_H^{2 n + 2}}{r^{2 n + 4}}
= - \frac{\sigma_n / (n+2)^2}{z^2}.
\end{equation}
Thus, the wave equation 
\begin{equation}
\label{schro3}
\frac{d^2 \Psi}{d z^2} + \left(\omega^2 - V_{\mbox{eff}}(r) \right)
\Psi = 0
\end{equation}
simply provides a solution
\begin{equation}
\label{solu1}
\Psi_{0,A} \sim \sqrt{2\pi \omega z}
\left[A_+ J_{\nu} (\omega z) + A_- J_{-\nu} (\omega z) \right]
\end{equation}
in this regime where
\begin{equation}
\label{argu1}
\nu = \sqrt{\frac{1}{4} - \frac{\sigma_n}{(n+2)^2}}.
\end{equation}

Taking a $\omega z \rightarrow \infty$ limit in $\Psi_{0,A}$, we can obtain
one asymptotic solution
\begin{equation}
\label{solu2}
\Psi_{\infty,A} \sim 
\left[A_+ e^{i \pi (1 + 2 \nu) / 4} + A_- e^{i \pi (1 - 2 \nu) / 4} \right]
e^{-i \omega z}
\end{equation}
with a constraint
\begin{equation}
\label{constraint1}
A_+ e^{-i \pi (1 + 2 \nu)/4} + A_- e^{-i \pi (1 - 2 \nu) / 4} = 0.
\end{equation}

To follow the contour given in Ref.\cite{motl03} we should turn to an 
angle $3 \pi$ 
around $z = 0$ in $\Psi_{0,A}$. Then we can derive another solution 
in the $r \sim 0$
regime
\begin{equation}
\label{solu3}
\Psi_{0,B} \sim \sqrt{2 \pi \omega z e^{i 3 \pi}}
\left[A_+ J_{\nu} (\omega ze^{i 3 \pi}) + A_- J_{-\nu} (\omega ze^{i 3 \pi}) 
\right].
\end{equation}
Since $J_{\nu}(z) = z^{\nu} \varphi(z)$ where $\varphi(z)$ is an even
analytic function in the entire region of the complex plane, it is easy 
to show that $\Psi_{0,B}$ reduces to 
\begin{equation}
\label{solu3-1}
\Psi_{0,B} \sim e^{i 3 \pi / 2} \sqrt{2\pi \omega z}
\left[A_+ e^{i 3 \pi \nu} J_{\nu} (\omega z) + A_- e^{-i 3 \pi \nu}
J_{-\nu} (\omega z) \right].
\end{equation}
Taking $\omega z \rightarrow - \infty$ limit in $\Psi_{0,B}$, we can
obtain another asymptotic solution
\begin{equation}
\label{solu4}
\Psi_{\infty,B} \sim
\left[A_+ e^{i 5 \pi (1 + 2 \nu) / 4} + A_- e^{i 5 \pi (1 - 2 \nu)/4} \right]
e^{-i \omega z} + 
\left[A_+ e^{i 7 \pi (1 + 2 \nu) / 4} + A_- e^{i 7 \pi (1 - 2 \nu)/4} \right]
e^{i \omega z}.
\end{equation}

Now, we can follow the contour to return to the initial point through 
the asymptotic region. Since the second term in Eq.(\ref{solu4}) is exponential
small in this region, the monodromy can be written as
\begin{equation}
\label{monod2}
{\cal M}(1) = 
\frac{A_+ e^{i 5 \pi (1 + 2 \nu) / 4} + A_- e^{i 5 \pi (1 - 2 \nu)/4}}
     {A_+ e^{i \pi (1 + 2 \nu) / 4} + A_- e^{i \pi (1 - 2 \nu)/4}}.
\end{equation}
With an aid of the constraint (\ref{constraint1}) this simply reduces to
\begin{equation}
\label{monod3}
{\cal M}(1) = - (1 + 2 \cos 2 \pi \nu).
\end{equation}
Thus equating Eq.(\ref{monod1}) with Eq.(\ref{monod3}) gives
\begin{equation}
\label{final}
e^{\omega / T_H} = - (1 + 2 \cos 2 \pi \nu).
\end{equation}

Now, we turn to the scalar field perturbation. In this case $\nu$ becomes 
\begin{equation}
\label{nus}
\nu_S = \sqrt{\frac{1}{4} - \frac{n+1}{(n+2)^2}}.
\end{equation}
Note that $\nu_S = 2/3$ when $n=4$, which makes the rhs of Eq.(\ref{final})
to be zero. When $n < 4$, the asymptotic quasinormal frequencies become
\begin{equation}
\label{sqnf1}
\frac{\omega_S}{T_H} = 2 \pi i \left( \bar{n} + \frac{1}{2} \right) + 
\ln (1 + 2 \cos 2 \pi \nu ) \hspace{2.0cm} (\bar{n} = 0, 1, 2, \cdots).
\end{equation}
Of course, $\nu_S = 0$ when there is no extra dimension, which gives a 
well-known result $Re(\omega_S / T_H) = \ln 3$. When extra dimensions
exist, the real parts of the asymptotic quasinormal frequencies are 
summarized in Table I\footnote{In Ref.\cite{ber03-2} it is shown that
when $n=1$, the real part of the lowest quasinormal frequency is 
$0.273$ which is much larger than the asymptotic value $T_H \ln 2 = 0.055$}.

\vspace{1.0cm}
\begin{center}

\begin{tabular}{l|lr} \hline
$n$                &          $Re(\omega_S / T_H)$        \\  \hline \hline
$0$                &          $\ln 3 = 1.09861$         \\
$1$                &          $\ln 2 = 0.69315$         \\
$2$                &          $0$                        \\
$3$                &          $\ln ((3 - \sqrt{5}) / 2) = -0.96242$   \\
$4$                &          $-\infty$                   \\  \hline
\end{tabular}

\vspace{0.2cm}
\large{Table I} 
\end{center}
\vspace{1.0cm}

It is interesting to note that $Re(\omega / T_H)$ vanishes when there are 
two extra dimensions. When $n > 4$, the rhs of Eq.(\ref{final}) becomes 
positive. Thus the asymptotic quasinormal frequencies become
\begin{equation}
\label{sqnf2}
\frac{\omega_S}{T_H} = \ln (-1 - 2 \cos 2 \pi \nu_S)
\end{equation}
In this case, therefore, we get $Im(\omega_S / T_H) = 0$. If there are infinite
extra dimensions, $\nu_S$ becomes $1/2$, which makes $Re(\omega_S / T_H) = 0$.

In the gravitational perturbation $\nu$ becomes 
\begin{equation}
\label{nug}
\nu_G = \sqrt{\frac{1}{4} + \frac{(n+1)^2 + 2}{(n+2)^2}}.
\end{equation}
It is interesting to note that $\nu_G = 1$ when $n=0$ and $n=4$. Thus in 
addition to the case of no extra dimension we get a well-known result
$Re(\omega_G / T_H) = \ln 3$ when there are four extra dimensions. Unlike the 
case of the scalar field perturbation, however, the rhs of Eq.(\ref{final}) is 
always negative regardless of $n$. Thus, the asymptotic quasinormal frequencies
are always given by
\begin{equation}
\label{gqnf1}
\frac{\omega_G}{T_H} = 2 \pi i \left(\bar{n} + \frac{1}{2} \right) + 
\ln (1 + 2 \cos 2 \pi \nu_G ) \hspace{2.0cm} (\bar{n} = 0, 1, 2, \cdots).
\end{equation}
The real part of $\omega_G / T_H$ is plotted in Fig. 1.

\begin{figure}[ht!]
\begin{center}
\epsfysize=8.0cm
\epsfbox{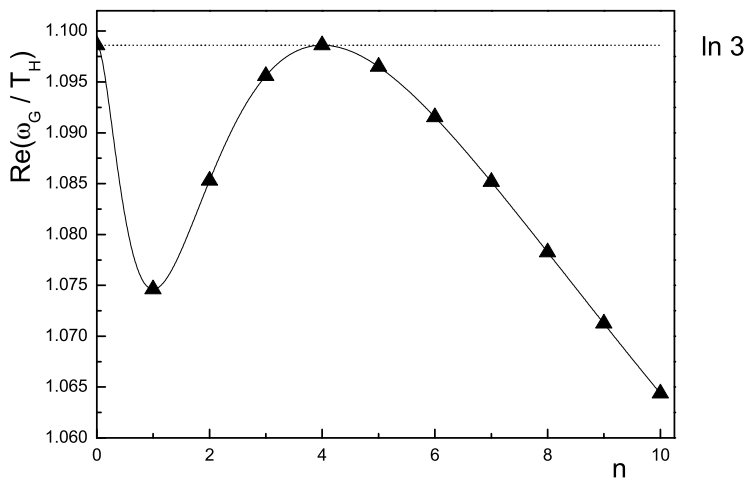}
\caption[fig1]{The $n$-dependence of $Re(\omega_G / T_H)$. It is interesting
to note that the well-known result $Re(\omega_G / T_H) = \ln 3$ is 
reproduced when $n=0$ and $n=4$. For other value of $n$ $Re(\omega_G / T_H)$
is shown to be smaller than $\ln 3$.}
\end{center}
\end{figure}

Fig. 1 shows that $Re(\omega_G / T_H)$ becomes the well-known result
$\ln 3$ when $n=0$ and $n=4$ as mentioned before. When $n$ is other 
number\footnote{When $n=1$, Ref.\cite{ber03-2} shows that the real part
of the lowest quasinormal frequency is $0.805$ which is much larger
than the asymptotoc value $0.086$.},
$Re(\omega_G / T_H)$ becomes smaller than $\ln 3$. In particular, if there
are infinite extra dimensions, $\nu_G$ becomes $\sqrt{5} / 2$, which gives 
$Re(\omega_G / T_H) = \ln (1 + 2 \cos \sqrt{5} \pi) \approx 0.906$.

In this letter we have computed the asymptotic quasinormal frequencies of the
brane-localized $(4+n)$-dimensional Schwarzschild black hole. For the case of 
the scalar field perturbation the well-known result $Re(\omega_S / T_H) = \ln 3$
is reproduced when there is no extra dimension. When $0 < n < 4$, 
$Re(\omega_S / T_H)$ is shown to be smaller than $\ln 3$ When $4 < n$, 
$Im(\omega_S / T_H)$ becomes zero.
In the case of the gravitational perturbation, however, $Im(\omega_G / T_H)$
does not vanish regardless of the number of the extra dimensions. 
In particular, the well-known result $Re(\omega_G / T_H) = \ln 3$ is 
obtained when $n=0$ and $n=4$. 

It is of interest to extend our computation to the brane-localized
$(4+n)$-dimensional Reissner-Nordstr\"{o}m and Kerr black holes. 
It is known that the asymptotic quasinormal frequencies of the $4d$ 
non-extremal
Reissner-Nordstr\"{o}m black hole have real part $T_H \ln 2$ unlike the
Schwarzschild black hole case\cite{motl03}. It is of interest to examine
how the existence of the extra dimensions changes the quasinormal frequencies. 
In the case of the Kerr black hole the superradiance effect 
seems to play an important role\cite{berti03-2}. 
Since it is proven generally that the superradiance modes exist 
in the brane-localized
and bulk rotating black holes\cite{ida03,jung05-3}, it seems to be interesting 
to examine how these modes
affect the asymptotic quasinormal spectrum.
It is also interesting 
to confirm our result by adopting an appropriate numerical technique. 

The most interesting issue at least for me is to explore how the $n$-dependent
quasinormal modes affect the Hawking radition of the black holes. Similar
issue was studied long ago by York\cite{york83}. We would like to study 
this issue in the future.
\vspace{1cm}

{\bf Acknowledgement}:  
This work was supported by the Korea Research
Foundation under Grant (KRF-2003-015-C00109).

\end{document}